\documentclass[acmsmall,nonacm]{acmart}
\startPage{1}
\setcopyright{none}

\usepackage{booktabs}
\usepackage{rotating} 

\definecolor{halfgray}{gray}{0.5}
\definecolor{deepblue}{rgb}{0,0,0.5}
\definecolor{deepred}{rgb}{0.6,0,0}
\definecolor{deepgreen}{rgb}{0,0.5,0}
\definecolor{highlightorange}{rgb}{0.98, 0.92, 0.8}

\usepackage{tikz}
\usetikzlibrary{calc}
\usetikzlibrary{shapes,arrows}
\usetikzlibrary{positioning}

\makeatletter
\newenvironment{btHighlight}[1][]
{\begingroup\tikzset{bt@Highlight@par/.style={#1}}\begin{lrbox}{\@tempboxa}}
{\end{lrbox}\bt@HL@box[bt@Highlight@par]{\@tempboxa}\endgroup}

\newcommand\btHL[1][]{%
  \begin{btHighlight}[#1]\bgroup\aftergroup\bt@HL@endenv%
}
\def\bt@HL@endenv{%
  \end{btHighlight}%
  \egroup
}
\newcommand{\bt@HL@box}[2][]{%
  \tikz[#1]{%
    \pgfpathrectangle{\pgfpoint{1pt}{0pt}}{\pgfpoint{\wd #2}{\ht #2}}%
    \pgfusepath{use as bounding box}%
    \node[anchor=base west, fill=highlightorange, outer sep=-1pt, inner xsep=2.5pt, inner ysep=.3pt, rounded corners=1pt, minimum height=\ht\strutbox+1pt,#1]{\raisebox{1pt}{\strut}\strut\usebox{#2}};
  }%
}
\makeatother

\hypersetup{
    colorlinks, 
    linkcolor={deepred},
    citecolor={deepgreen},
    urlcolor={deepblue}
}

\usepackage[linesnumbered,ruled,noend]{algorithm2e}

\def\Snospace~{\S{}}

\usepackage{listings}

\lstdefinelanguage{JavaScript}{
  morekeywords=[1]{break, continue, delete, else, for, function, if, in,
    new, return, this, typeof, var, let, const, void, while, with},
  morekeywords=[2]{false, null, true, boolean, number, undefined,
    Array, Boolean, Date, Math, Number, String, Object},
  morekeywords=[3]{eval, parseInt, parseFloat, escape, unescape},
  sensitive,
  morecomment=[s]{/*}{*/},
  morecomment=[l]//,
  morecomment=[s]{/**}{*/}, 
  morestring=[b]',
  morestring=[b]"
}[keywords, comments, strings]

\makeatletter
\lst@AddToHook{PreSet}{\normallineskiplimit=0pt}
\makeatother

\usepackage{cleveref} 

\usepackage[graphicx]{realboxes}
\usepackage{csquotes} 

\newcommand{\circled}[1]{\kern-1pt\raisebox{.4pt}{\textcircled{\raisebox{-.4pt}{\kern.2pt\footnotesize#1}}}}

\usepackage{enumitem} 
\usepackage{calc}
\usepackage{tabularx}

\begin{document}

\title{Systematic Generation of Conformance Tests for JavaScript}

\author{Blake Loring}
\affiliation{
  \position{}
  \department{Information Security Group}
  \institution{Royal Holloway, University of London}
  \country{United Kingdom}
}
\email{blake.loring.2015@rhul.ac.uk}

\author{Johannes Kinder}
\affiliation{
  \position{}
  \department{Research Institute CODE}
  \institution{Bundeswehr University Munich}
  \country{Germany}
}
\email{johannes.kinder@unibw.de}

\begin{abstract}
  JavaScript implementations are tested for conformance to the ECMAScript
  standard using a large hand-written test suite. Not only in this a tedious
  approach, it also relies solely on the natural language specification for
  differentiating behaviors, while hidden implementation details can also affect
  behavior and introduce divergences. We propose to generate conformance tests
  through dynamic symbolic execution of polyfills, drop-in replacements for
  newer JavaScript language features that are not yet widely supported. We then
  run these generated tests against multiple implementations of JavaScript,
  using a majority vote to identify the correct behavior.
  To facilitate test generation for polyfill code, we introduce a model for
  structured symbolic inputs that is suited to the dynamic nature of
  JavaScript. In our evaluation, we found 17 divergences in the widely used
  core-js polyfill and were able to increase branch coverage in interpreter code
  by up to 15\%. Because polyfills are typically written even before
  standardization, our approach will allow to maintain and extend
  standardization test suites with reduced effort.
\end{abstract}

\maketitle

\section{Introduction}

JavaScript started out as a lightweight scripting language for the web, but has
grown to become a language powering large web applications, server-side
backends, and embedded systems. As a result of its rushed initial development
and the following rapid growth, it is also a complex and sometimes quirky
language.
Its growing importance as an industry standard with several competing
implementations has led to it being standardized as ECMAScript, with a precise
and highly complex specification document.
As the language evolved, it has become common for not all implementations to
support the latest language features and APIs. To make use of them but retain
compatibility, developers can use \textit{polyfills}, pure JavaScript libraries
that simulate new language features if they are not yet supported by the
interpreter.

Smooth interoperability between interpreters by different vendors and the
various polyfills is ensured by the common ECMAScript standard and its test
suite.  While there is a formally verified reference interpreter for the core
language, which closely follows the natural language
specification~\cite{BodinCFGMNSS14}, all fully-fledged implementations in
browsers and other systems rely on test suites to ensure conformance.
The main mechanism for validating conformance to the ECMAScript standard is
Test262~\cite{ecma_test262}, a manually curated test suite with the goal of
covering all observable behavior of the ECMAScript specification.

Because Test262 is created manually, it is likely that it does not entirely
achieve this goal. Furthermore, implementations of a specification by definition
add additional implementation detail. As a consequence, we argue that
interpreters contain relevant behavior that is not exercised by Test262. When
corner cases remain untested, there is a potential for hidden divergences from
the specification.
Methods from automated test generation seem ideally suited to fill this gap and
exercise hidden behavior. Techniques such as dynamic symbolic execution promise
to generate high-coverage test suites fully automatically with the help of an
satisfiability-modulo-theories (SMT) solver. In principle, such techniques will
allow to generate test cases for \textit{implementations} of ECMAScript language
semantics.
Full JavaScript implementations are highly complex software systems,
however. There has been some success using simpler testing techniques such as
fuzzing to find bugs in interpreters~\cite{langfuzz}, and also using dynamic
symbolic execution to test interpreters for simpler
languages~\cite{chef-asplos14}. But so far, dynamic symbolic execution does not
scale to full JavaScript interpreters, and interpreter features such as
just-in-time compilation make full support highly unlikely.

In this paper, we propose to symbolically execute straightforward
implementations of JavaScript language features to generate new test
cases. These are then executed on a portfolio of JavaScript interpreters, using
a \textit{majority vote} to decide the correct behavior.  We find that polyfills
are ideally suited for this task: polyfills are directly executable and provide
more detail than the ECMAScript specification; at the same time, they are much
more compact than implementations in an interpreter. Polyfills also have the
advantage of providing a clear entry point for each supported feature, which
makes directed testing possible. In interpreters, the implementation of language
semantics is hidden behind parsing and translation layers, far removed from any
external entry point that could be controlled by a test generation tool.

Entry points of polyfill code can require structured input such as objects and
arrays, whereas dynamic symbolic execution usually only yields primitive input
values returned by the SMT solver.
Due to the lack of static typing in JavaScript, object and array creation cannot
rely on type information as is usually done for object-oriented
languages~\cite{symstra,test_input_generation_jpf}. Instead, we introduce a
purely dynamic approach to generate structured test inputs in dynamic symbolic
execution. We intercept accesses to object fields and array elements by the
JavaScript program at runtime and generate test cases for each meaningful
outcome, handling possible name aliasing, different field types, and the
special quirks of JavaScript arrays.

We evaluate our approach in an implementation on top of
ExpoSE~\cite{expose-spin17}, an existing dynamic symbolic execution engine for
JavaScript programs. We automatically generate a rich suite
of tests from the Mozilla Developer Network polyfills (\lstinline|mdn-polyfills|)
and \lstinline|core-js| that we run against SpiderMonkey, Node.js, and QuickJS. 
In summary, we make the following contributions:
\begin{itemize}
\item We present a methodology for automated generation of conformance tests
  from polyfills. We employ differential testing across multiple implementations
  to compensate the lack of testing oracles (\autoref{sec:conformance_testing}).
\item We define a model for symbolic objects and symbolic arrays that
  dynamically synthesizes test inputs in untyped JavaScript code
  (\autoref{sec:symbolic_datastructures}).
\item We improve the state of the art in conformance testing of ECMAScript
  implementations through our methodology. Using our new tests, we found 17 bugs in
  polyfill implementations and were able to augment the coverage of Test262 in
  JavaScript interpreters by up to 15\%~(\autoref{sec:eval}).
\end{itemize}

Overall, we believe that this can lower the bar for maintaining standardization
test suites like Test262 in the future. New language features are regularly
implemented in polyfills before standardization, and our approach will allow to
generate corresponding tests as a byproduct.

\section{Background}

We begin by providing the necessary background on
JavaScript~(\autoref{sec:js_values}), dynamic symbolic execution~(\autoref{sec:dse}),
and the particulars of applying it to test JavaScript code~(\autoref{sec:dsejs}).

\subsection{JavaScript}
\label{sec:js_values}

JavaScript has a dynamic type system, so program source code contains no type
annotations and no type checking is performed during preprocessing. Instead,
type information (tags) is attached to values when they are created and rules are
enforced at runtime.
All values are in the same form, a structure consisting of data and a tag
that indicates the value type.
Whenever the interpreter executes an instruction, it first inspects the type
of operands. If the operands are not in the desired type, then a series of
type coercion rules are applied to convert them to usable types or a type error is thrown.
For example, \lstinline|"Hello " + 5| results in the string \lstinline|"Hello 5"|.
Automatic type coercions can have unintuitive semantics. For example, in
contrast to the previous expression, \lstinline|"Hello" - 5| evaluates to
\lstinline|NaN|, since a string added to a number coerces the number to a
string, but a string subtracted from a number attempts to coerce the string
to a number.
The combination of dynamic typing and automatic type coercion can make
bugs in programs hard to track, since applying operations on incompatible
types will not cause an immediate error and instead propagate through
the program. For example, the following program will produce the result
\lstinline|"NaNHello10"|. If y is not fixed, it will be difficult to
identify where first error occurs:
\begin{lstlisting}
function doTask(y) {
  let j = y + 10;
  let q = 4 - y + j;
  return q;
}
doTask('Hello');
\end{lstlisting}

Objects are maps from string property names to values. Objects
are constructed dynamically and have no pre-set structure.
\lstinline|let x = { a: 'H' };| constructs a new object with a single
property, \lstinline|a| and assigns it to the variable \lstinline|x|. After
executing \lstinline|x.b = 'Q'|, x will have two properties set, \lstinline|a|,
and \lstinline|b|.
Values of any type can be assigned to object properties, including other objects,
arrays, and functions. For example, the following code will create a new
function and assign it to the property \lstinline|printA| on \lstinline|x|:
\begin{lstlisting}
x.printA = function() {
  console.log(this.a);
}
\end{lstlisting}
When a function attached to an object is executed, the containing object is
passed as the \lstinline|this| argument to the function call.
For example, \lstinline|x.printA();| will print \lstinline|H|, since
\lstinline|this| refers to \lstinline|x|.

The language also allows object-oriented programming.
Classes are constructed dynamically through constructor functions and the
\lstinline|new| keyword.
When \lstinline|new| is used, a fresh object will be created and the
constructor is executed with the \lstinline|this| value equal to the new
object. The resulting object is returned after construction.
For example, the following code defines a new class and then creates an
instance of it with the argument \lstinline|"Hello"|:
\begin{lstlisting}
function A(arg) {
	this.arg = arg;
}
let a = new A("Hello");
\end{lstlisting} 
Note that there is no distinction between class constructors and other
functions.
In our example, since \lstinline|A| is a function, we are also allowed
to execute it without the \lstinline|new| keyword.
This allows created objects to call other class constructors to simulate inheritance.
For example, the following code will create a class constructor \lstinline|B|, and use the \lstinline|A| constructor to make sure it has the same properties:
\begin{lstlisting}
function B(arg) {
  A.call(this, arg);
}
\end{lstlisting}

Assigning all properties of a new object in the constructor can make
managing code difficult, so the language also includes object prototypes.
If we want a value to be added to every instance of \lstinline|A|, then we
can add it to the \lstinline|prototype| object which exists as a property
of every function.
For example, once we execute \lstinline|A.prototype.q = 'bye'|, the property
\lstinline|q| will exist in any new instance of A.
These prototypes can be chained together, forming inheritance chains.
For example, the following code defines \lstinline|B| an extension of
\lstinline|A|, and will print \lstinline|bye| since it inherits \lstinline|q|
from the chained prototype:
\begin{lstlisting}
function B() {
	B.prototype.constructor.call(this, "Hello");
}
B.prototype = Object.create(A.prototype);
console.log((new B()).q);
\end{lstlisting}
Prototype chaining and prototypal inheritance are core to object abstractions in JavaScript.
While later revisions to the standard add support for the \lstinline|class|
keyword, this is just syntactic sugar for prototypes.

\subsection{Dynamic Symbolic Execution}
\label{sec:dse}

Dynamic symbolic execution (DSE) is an automated test generation
approach based on constraint solving and has been shown to be effective at
bug-finding~\cite{klee,GodefroidLM08,BounimovaGM13}.
DSE generates new test cases for a program through repeat executions. In DSE,
some inputs to a program as marked as symbolic while others are fixed. The
DSE engine then generates a series of assignments for symbolic values which
each exercise a unique control flow path through the program.
For example, when analyzing the following program, we begin by replacing
the input \lstinline|x| with the symbol $X$:
\lstset{escapeinside={*@}{@*}}
\begin{lstlisting}[escapechar=§]
var x = §$X$§;
§\label{dse_example_y}§var y = x + x;
§\label{dse_example_if}§if (x > 10) {
§\label{dse_example_infeasible_if}§  if (y < 20) {
    ...
  }
}
\end{lstlisting}

When executing the program we maintain a symbolic state in addition to the
concrete state. The concrete state drives test execution, while the symbolic
state tracks constraints on the symbols in the program.
To begin analysis, we execute the test harness with an initial concrete
assignment for the symbolic inputs.
For our example, we pick the initial assignment \lstinline|x = 5|.

With our test setup and our initial test case selected, we are now ready to symbolically execute the program.
When operations involve symbolic operands, we compute the concrete result
using the concrete value of the operand and use a symbolic interpreter to
generate the resulting symbol.
We see this on line \ref{dse_example_y}, where execution with our initial
test case will yield a concrete value of \lstinline|y = 10|, and a symbolic
value of \lstinline|y = $X$ + $X$|.
We now reach line \ref{dse_example_if}, the first branching condition in the program. 
In DSE, we use use the concrete state to decide which branch to follow for
the current test execution and we also develop a symbolic path condition
(PC), a symbolic representation of the conditional operations which drove
us down the branches we followed.
On line \ref{dse_example_if} we follow the \lstinline|else| branch, and
do not enter the if condition since the concrete value of \lstinline|x| is
\lstinline|5|.
We use the $\leftarrow$ operator to denote updates to the symbolic path condition. At this step we update our path condition with $PC \leftarrow PC \land X < 10$.
After this, our first test case terminates.

Upon termination, the DSE engine uses the PC and an SMT solver in order to
find alternate assignments for the symbolic inputs.
We find these alternate assignments by negating the conditional operations
in the PC so that the next test case will take the opposite route at that
branching point.
We now try and find an alternative assignment for $X$ which will follow the
\lstinline|true| branch on line \ref{dse_example_if}.
We query the SMT solver to decide there is any assignment for $X$ where $X >
10$, and the SMT solver gives us the input $X = 25$, our new test case.

Since we have identified a new test case, we now re-execute our program with the new concrete assignment for $X$.
During this execution we follow the \lstinline|true| path on line \ref{dse_example_if}, and each line \ref{dse_example_infeasible_if} with the path constraints $PC = X > 20$.
On line \ref{dse_example_infeasible_if}, we check if \lstinline|y < 20|.
In this test case, \lstinline|y| has a concrete value of 50, and a symbolic value of $X + X$.
Since 50 is greater than 20, we take the else path and update the PC with
$PC \leftarrow PC \land X + X > 20$, leading to test case termination.

We now use the SMT solver to decide if there is an assignment for $X$ which
explores the true branch on line \ref{dse_example_infeasible_if}.
We take the PC and negate the last constraint, resulting in a query asking
the SMT solver if there is a feasible assignment for X such that $X > 10
\land X + X < 20$.
Here, the SMT solver tells us that there is no feasible assignment for
$X$, so we know that the true branch on line \ref{dse_example_infeasible_if}
is unreachable.
Since there are no new test cases for our program our DSE is now complete
and we have explored all feasible control flows contingent on our symbol $X$.
In general, there will be an impractical (possibly infinite) number of test cases to execute.
So instead of exhausting all test cases, we repeatedly execute new test
cases until we reach a time limit or a predefined coverage goal.
Therefore, DSE can in general not be used for software verification, but it is
ideally suited to generate high-coverage test suites fully automatically.

\subsection{Dynamic Symbolic Execution for JavaScript}
\label{sec:dsejs}

The complex dynamic type system, dynamic nature of programs, and rapid pace
of change in the language make JavaScript programs challenging to symbolically execute.
Additionally, programs use a lot of high level features, such as objects,
arrays, strings and regular expressions which can be tricky to reason about
symbolically.

\citet{SaxenaAHMMS10}, \citet{symjs:Li}, and \citet{javert_es5} built custom
symbolic interpreters for JavaScript, but the language changes frequently and
these engines target older versions of the standard, making them impractical for
current real-world analysis.
\citet{jalangi} took an alternate approach when developing Jalangi, a symbolic
framework which uses program instrumentation to embed the symbolic engine
directly into a program. By instrumenting the program source code maintenance
cost is reduced, but it is harder to segregate the symbolic state from the
running program.
Jalangi does not support symbolic regular expressions, and only includes
a limited support for strings. The engine is also no longer supported,
but can still be run on ES5 programs.

To generate our conformance tests we use ExpoSE. ExpoSE is a open-source
DSE engine for modern JavaScript~\cite{expose-spin17} designed for practical
symbolic execution.
The engine separates test case scheduling, SMT solving, and test execution
which allows for concurrent executions.
In ExpoSE, test executions are isolated to avoid artifacts from asynchronous events
impacting subsequent executions.
ExpoSE uses Jalangi2~\cite{Jalangi2} to instrument programs,
embedding the symbolic execution engine into the code.
To propagate symbolic values, ExpoSE uses \emph{concolic} values, where a
symbolic value includes both a symbolic expression and a concrete value for
that test case. These values are propagated through the program instead of
standard JavaScript values.
When performing operations, the instrumented program will first check if
operands are symbolic. When symbolic, the instrumentation
will call a symbolic interpreter to develop the symbolic expression before
directly evaluating the concretely portion.
ExpoSE uses the Z3 constraint solver to find alternate test cases,
and includes support for strings and ES6 regular expressions out of the
box~\cite{pldi19-regex}.

\paragraph{Modifications to ExpoSE}

In addition to adding support for symbolic
objects~(\autoref{ssec:symbolic_object_full}), we made additions to ExpoSE so
that it can treat type-coercions we observed in existing polyfills.
In JavaScript, numeric values may be either integers or floating point values
and there is no idiomatic way to ensure that a value is an integer.
If a developer wants to force a number to be an integer they often use
a bitwise operation to force the coercion, since bitwise logic truncates
operands to integers.
To illustrate this, the \lstinline{targetLength = targetLength >> 0} ensures
that the length is an integer with a bitshift by 0.
ExpoSE did not accurately model bitwise operations and other esoteric
behaviors of the type system, but these are used often in built-in
implementations, so we modified the engine to support them.

\section{Conformance Testing using PolyFills}
\label{sec:conformance_testing}

We generate new implementation conformance tests for JavaScript interpreters
through symbolic execution of polyfills; implementations of built-in methods
in JavaScript.
Existing supplementary test suites like Test262, the official ECMA test
suite~\cite{ecma_test262}, are created by exploring conditions in the
specification.
Since they are manually curated, bugs may be missed -- particularly when
treating edge cases.
Here, we use a DSE engine to automatically explore the subtleties of built-in
implementations in polyfills, and then apply the generated tests to other
implementations, since they should all behave identically.

\subsection{Polyfills}

With each evolution of JavaScript there is a period of time where new feature
support will not be ubiquitous, since each vendor will take time to update
their implementation. To remedy this, polyfills, short programs
which implement built-in methods, have become common. A polyfill will
inject a built-in into the standard library at runtime if it is not already
supported by the host interpreter.
 
In this paper we use two polyfill packages, \lstinline|core-js|~\cite{core-js} and
\lstinline|mdn-polyfills|~\cite{MDNPolyfill}, to test our approach. These libraries contain
polyfill implementations of standard library methods added in the ES6
standard. \lstinline|core-js| is the de-facto standard for polyfills with
78,000,000 monthly downloads. \lstinline|mdn-polyfills| is less highly depended on,
with 72,000 monthly downloads on NPM, the largest JavaScript package repository.

\subsection{Architecture}

We generate new test cases by dynamic symbolic execution of polyfills.
Analysis of these polyfills will generate inputs that explore the intricacies of
built-in specifications, but we do not have a ground-truth for the correct
behavior of a test case.
To solve this problem we use a suite of interpreters and have
them vote on the correct answer.
This acts an oracle to identify when an implementation is incorrect, and
only requires manual intervention when two or more implementations diverge.

We split our implementation into two components, the test case generator,
and the test case executor. The test case generator uses ExpoSE to generate
new test cases.
The test case executor executes a test suite extracted from the symbolic
executions and checks that each of our selected interpreters is implemented
correctly.

\subsection{Test Case Generation}

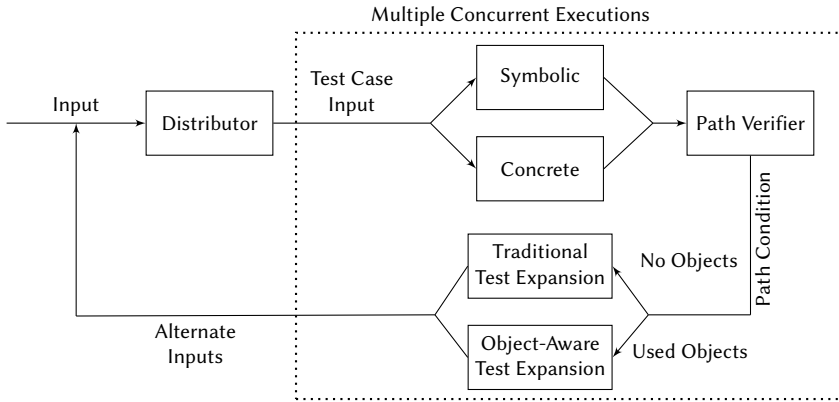
\begin{figure*}[t]%
\tikzstyle{block} = [draw, fill=white, rectangle, minimum height=3em, minimum width=6em]%
\tikzstyle{pinstyle} = [pin edge={to-,thin,black}]%
\begin{center}%
\begin{tikzpicture}[auto, node distance=2.25cm,>=latex',font=\sffamily,scale=0.8,every node/.style={scale=0.8}]

  \node [align=center] (seedinputs) {};
  \node [block, right of=seedinputs, node distance=3.5cm] (distributor) {Distributor};
  \node [right of=distributor, node distance=3.8cm] (distributor fork) {};

  \node [block, right of=distributor, node distance=5.5cm, yshift=-0.75cm] (oracle) {Concrete};
  \node [block, above of=oracle, align=center, node distance=1.5cm] (testcase) {Symbolic};

  \node [block, below of=oracle, align=center, node distance=1.6cm] (traditional) {Traditional\\Test Expansion};
  \node [block, below of=traditional, align=center, node distance=1.5cm] (objectaware) {Object-Aware\\Test Expansion};
  \node [right of=traditional, node distance=3.5cm, yshift=-0.8cm] (expansion fork) {};
  \node [left of=expansion fork, node distance=1.7cm] (expansion merge) {};
  \node [left of=traditional, node distance=1.75cm, yshift=-0.8cm] (expansion fork 2) {};

  \node [block, right of=oracle, node distance=3.5cm, yshift=0.75cm] (results) {Path Verifier};
  \node [right of=results, align=center, node distance=3.5cm] (testsuite) {};
  \node [left of=results, node distance=1.5cm] (results fork) {};

  \draw [->] (seedinputs.east) -- node[name=inputnode] {Input} (distributor.west);

  \draw [-] (distributor.east) -- node[name=u, align=center] {Test Case\\Input} (distributor fork.west);
  \draw [->] (distributor fork.west) -- (testcase.west);
  \draw [->] (distributor fork.west) -- (oracle.west);

  \draw [-] (testcase.east) -- (results fork.west);
  \draw [-] (oracle.east) -- (results fork.west);
  \draw [->] (results fork.west) -- (results.west);

  \draw [-] (results.south) -- node[name=u, yshift=-1.3cm, xshift=0.2cm, rotate=90] {Path Condition} (expansion fork.center);
  \draw [-] (expansion fork.center) -- (expansion merge.center);
  \draw [->] (expansion merge.center) -- node[name=u, xshift=1.9cm, yshift=0.75cm] {No Objects} (traditional.east);
  \draw [->] (expansion merge.center) -- node[name=u, xshift=-0.1cm] {Used Objects} (objectaware.east);

  \draw [-] (traditional.west) -- (expansion fork 2.center);
  \draw [-] (objectaware.west) -- (expansion fork 2.center);

  \node [below of=inputnode, node distance=3.45cm] (input join) {};
  \draw [-] (expansion fork 2.center) -- node[name=u, align=center, xshift=-1cm] {Alternate\\Inputs} (input join.center);
  \draw [->] (input join.center) -- (inputnode);

  \path (results.east) -- (results.east) coordinate (r1);
  \path (testsuite.west) -- (testsuite.west) coordinate (t1);

  \draw[thick,dotted] ($(testcase.north west)+(-3,0.2)$) rectangle ($(results.south east)+(0.5,-4)$); 
  \node[draw=none, fill=none, node distance=1cm, above of=testcase, align=left, xshift=-0.5cm, yshift=0.0cm] (concurrent){Multiple Concurrent Executions};
\end{tikzpicture}%
\end{center}%
\caption{Test Case Generator Overview.}%
\label{fig:overview_test_case_generator}%
\end{figure*}

We generate new test cases by symbolically executing polyfills using ExpoSE. 
\autoref{fig:overview_test_case_generator} provides an overview
of the architecture.
We begin by supplying the test apparatus with a target
built-in and the number of arguments the method expects.
The apparatus then constructs a series of symbolic inputs to use as arguments,
including a symbolic value for \lstinline|this|.
ExpoSE then analyzes the generated test harness and
begins to output a series of test cases.
We also execute each new test case in \lstinline|Node.js| to mitigate any
errors in ExpoSE.
We forward the result of the concrete and symbolic executions to the path
verifier, a tool that double-checks that the concrete and the symbolic result
are identical.
If they are not, then the test case is discarded. Otherwise, we add the test
case to the generated test case suite, and the symbolic path condition is
used to generate new test cases.
We use an object-aware type encoding when finding alternate test cases to
explore more of our target polyfills~(\autoref{sec:symbolic_datastructures}).

\subsection{Test Case Executor}
\label{ssec:test_executor}

\begin{figure}[t]%
\tikzstyle{block} = [draw, fill=white, rectangle, minimum height=3em, minimum width=6em]%
\tikzstyle{pinstyle} = [pin edge={to-,thin,black}]%
\begin{center}%
\begin{tikzpicture}[auto, node distance=2.25cm,>=latex',font=\sffamily,scale=0.8,every node/.style={scale=0.8}]
 \node [align=center] (seedinputs) {\\\\Input};
 \node [block, right of=seedinputs, node distance=1.5cm, align=center] (generator) {Test Case\\Generator}; 
 \node [block, right of=generator, node distance=3cm, align=center] (cjs) {Node.js (V8)};
 \node [block, above of=cjs, node distance=1.5cm, align=center] (chrome) {SpiderMonkey};
 \node [block, below of=cjs, node distance=1.5cm, align=center] (mdn) {\ldots};

 \node [block, right of=cjs, node distance=3cm, align=center] (voting) {Voting};
 \node [align=center, right of=voting, node distance=1.5cm] (outcome) {};

 \draw[thick,dotted] ($(chrome.north west)+(-0.4,0.7)$) rectangle ($(mdn.south east)+(0.4,-0.7)$); 

 \draw [->] (seedinputs.west) -- (generator.west);
 \draw [->] (generator.east) -- (cjs.west);
 \draw [->] (generator.east) -- (chrome.west);
 \draw [->] (generator.east) -- (mdn.west);

 \draw [->] (cjs.east) -- (voting.west);
 \draw [->] (chrome.east) -- (voting.west);
 \draw [->] (mdn.east) -- (voting.west);

 \draw [->] (voting.east) -- (outcome.east);

\end{tikzpicture}%
\end{center}%
\caption{Test Case Executor Overview.}%
\label{fig:test_executor_overview}%
\end{figure}
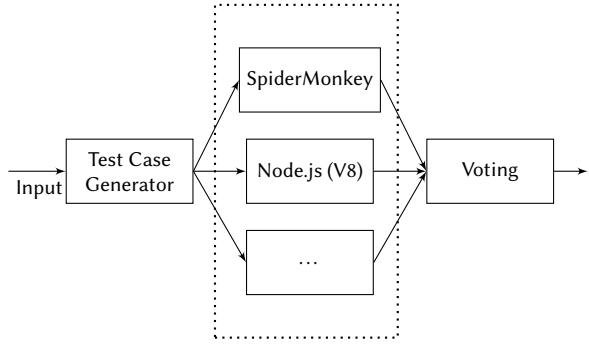

The second component in our design is the test case executor.
Our automatically generated test cases do not have a predetermined expected
result because the result found during symbolic execution may be from a
flawed implementation.
Instead of using predetermined test case results, we use a consensus-based
approach to detect incorrect implementations, illustrated in
\autoref{fig:test_executor_overview}.
We execute each test case in several different interpreters.
Each interpreter has a different interface so we generate a compatible test
through a test translator that takes a test input and returns a program
compatible with a specific engine.
For polyfills, we inject the target method into a Node.js instance, replacing
any existing implementation. We then execute each of these programs and
collect the output.

Once the test case has been executed by each implementation, we pass the
results to a voting mechanism.
The voting mechanism looks for implementations where behavior diverges from
the others. If the outcome of a built-in call diverges either in exception
type or result then we say that the interpreters disagree and raise an error.
Specifically, we say that an implementation disagrees if either of the following two conditions are violated:
\begin{enumerate}
  \item If a test case throws an exception and others do not, or the exception type
differs from other implementations.
  \item If a test case has output different from the others.
\end{enumerate}
We do not compare the exact text of exceptions because it is not
specified by the ECMAScript specification.
If a single implementation disagrees then it is marked as incorrect. When
multiple implementations disagree, we cannot make any conclusion about correct
behavior and mark the test case for manual review.

\section{Representing Symbolic Data Structures in JavaScript}
\label{sec:symbolic_datastructures}

To allow automated generation of structured test inputs for built-in methods, we
require a method for maintaining symbolic objects and arrays. We developed new
encodings for untyped symbolic objects, i.e., symbolic objects with no
pre-specified property names or types~(\autoref{ssec:symbolic_object_full}),
arrays of mixed types~(\autoref{sec:mixed_type_arrays}), and for homogeneously
typed arrays~(\autoref{sec:hom_arrays}).

\subsection{Motivation}

Support for symbolic objects is key to the exploration of built-ins because
it allows thorough exploration of object and array-centric built-ins.
More subtly, support allows the DSE engine to consider esoteric type-checking
in built-in methods.
The specification includes precise but unintuitive rules on how input values
are to be interpreted and when type contract violations should raise an error.
To highlight how an object encoding can improve coverage of these edge cases,
we now consider \lstinline|Array.prototype.find|.

Usually, this method is given an array as its base argument and a
predicate. The array is then searched, left to right, until a value satisfying
the predicate is found.
If no values satisfy the predicate then \lstinline|undefined| is returned.
For example, \lstinline|[11,23,20].find((x) => x % 2 == 0)| would yield \lstinline|20|, the first even number in the array.
If we look at the method specification, we see that there is a quirk to this
method contract. The method accepts any object which looks like an array
(i.e., any object with a length property).
Because of this \begin{sloppy}\lstinline[breaklines=true]|Array.prototype.find.call({0: 11, 1: 23, 2: 20, length: 3}, (x) => x % 2 == 0)|\end{sloppy} behaves equivalently to the previous example, but \begin{sloppy}\lstinline[breaklines=true]|Array.prototype.find.call({0: 11, 1: 23, 2: 20}, (x) => x 

One further quirk is the coercion of \lstinline|length| to an integer. The specification does not reject non-integer length properties, leading to a coercion \begin{sloppy} that resolves \lstinline[breaklines=true]|Array.prototype.find.call({0: 20, length: true}, (x) => x % 2 == 0)|\end{sloppy} to \lstinline|20|, but \begin{sloppy}\lstinline[breaklines=true]|Array.prototype.find.call({0: 20, length: false}, (x) => x 

In \lstinline|mdn-polyfills| these checks are implemented by \lstinline|var o = Object(this)|, which ensures the value is either an array or
an object, followed by \lstinline|var len = o.length >>> 0|, which
selects the length of the object and ensures it is an integer using type
coercion~\footnote{https://github.com/msn0/mdn-polyfills/blob/master/src/Array.prototype.findIndex/findIndex.js}.
In this case, if the length property is not an integer then it is first
coerced to a number and subsequently truncated to an integer.
Through our encoding of objects, we can synthesize useful test cases for
such behavior.

\subsection{Symbolic Objects}
\label{ssec:symbolic_object_full}

Representing JavaScript objects in DSE engines is challenging due to the dynamic
type system. Existing SMT solvers do not support a ``theory of
objects.'' Recreating a dynamic datatype in SMT and implementing the required
reasoning would complicate solver-side logic and effectively move much
language-specific reasoning into the SMT solver, which is designed to be
language-agnostic. So instead we opt to translate the reasoning about symbolic
objects into a form that can be represented as an SMT problem over primitive
types.
We develop an intermediate encoder that outputs typed SMT problems directly in
the DSE engine.
The intermediate encoder does not require solver extensions, instead
simulating symbolic objects by following every object operation
along a program trace and exploring feasible alternatives.

We model symbolic objects by tracking property lookups and
updates to objects.
For this, we rewrite all property lookups to use the
common interface \lstinline|getProperty(object, propertyName)| and all property
updates to use \lstinline|setProperty(object, propertyName, value)|.
In both cases, \lstinline|object| is the object
operated on and \lstinline|propertyName| is a string indicating which property is being accessed.
For \lstinline|setProperty|, \lstinline|value| is 
the new value of the given property.
With this instrumentation, we can keep track of all object operations during
execution, updating the symbolic state when appropriate.
We instrument arrays similarly, with \lstinline|getProperty| and \lstinline|setProperty|
interfaces for all property lookups. They differ in the typing of property names,
where they also accept integer values, since arrays can contain
integer and string property names.
 
\begin{figure}
\includegraphics[scale=0.2]{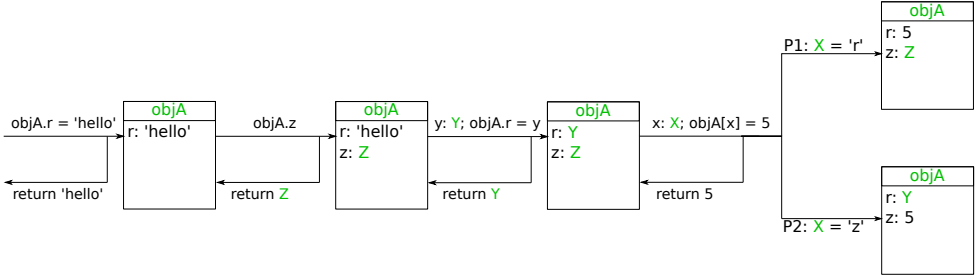}
\caption{Illustration of symbolic object modeling.}
\label{fig:expansion_explained}
\end{figure}

The root of our encoding is the creation of new symbolic values for properties
we have not seen before while returning the value stored in a state for
properties that we have previously set. Our encoding for objects is
illustrated in \autoref{fig:expansion_explained}. Here we see how a symbolic
object behaves under various typical operations.

The first step in \autoref{fig:expansion_explained} shows how symbolic
objects support fully concrete operations. Here, we record the concrete
value supplied to be returned on subsequent lookups.
When we perform a lookup for a property that we have not encountered before, we
introduce a new symbolic value to the program and set it to the appropriate
property.
The created symbol does not have a fixed type, and instead uses existing
support in the DSE engine to explore the program as if it were any of the
supported symbolic types.
In the case of ExpoSE, the DSE engine we use in this paper, the symbolic
types supported are undefined, null, boolean, number, string and through
our encoding also objects and arrays. The second operation illustrates this
process on \lstinline|objA| in \autoref{fig:expansion_explained}, where the
new symbol Z is introduced and assigned to the property z.

Next, we want to set a property with a concrete property name but a symbolic value.
As with a fully concrete set property, we record the supplied property value to
the object state; here, it makes no difference if the supplied properties are
concrete or symbolic.

The last matter that we address in this example is how we approach setting
and getting of properties with symbolic property names. Here, we attempt to create
new test cases for each of the previously recorded properties of an object -
even if they are subsequently deleted. The final operation illustrates this
in the figure, where we write a concrete value with a symbolic property name,
leading to two new paths. One where the property \lstinline|r| is replaced with 5,
and another where the property \lstinline|z| is replaced with \lstinline|5|. This
final step causes under-approximation in our encoding: We do not enumerate
on properties that we have not seen previously. We could, in principle, support
this through the enumeration of all possible property names, but this would
lead to an infeasible number of paths to explore.

\begin{algorithm}[t]
 \If{property is symbolic}{
  \For{knownProp in base} {
   \tcp{Attempt to generate a test case for each known property}
   \If{Concrete(property) $=$ knownProp} {
    $PC \leftarrow PC \land property = knownProp$\;
   } \Else { 
    $PC \leftarrow PC \land property \neq knownProp$\;
   }
  }
  \Return{base[property]};
 } \Else {
  \If {property not in base} {
   $base[property] = $$fresh$ $symbol$\;
  }
  \Return{base[property]}\;
 }
 \caption{Symbolic object encoder -- getProperty(base, property).}
 \label{alg:sym_encoder_getProperty}
\end{algorithm}

\begin{algorithm}[t]
 \If{property is symbolic}{
  \For{knownProp in base}{
   \tcp{Attempt to generate a test case for each known property}
   \If{Concrete(property) $=$ knownProp} {
    $PC \leftarrow PC \land property = knownProp$\;
   } \Else { 
    $PC \leftarrow PC \land property \neq knownProp$\;
   }
  }
  \Return{base[property] = value}\;
 } \Else {
  \tcp{Record the new value for property}
  \Return{base[property] = value}\;
 }
 \caption{Symbolic object encoder -- setProperty(base, property, value).}
 \label{alg:sym_encoder_setProperty}
\end{algorithm}

To implement our encoding we instrument the \lstinline|getProperty|
and \lstinline|setProperty| operations executed by a program with our
object encoder, detailed in \autoref{alg:sym_encoder_getProperty} and
\autoref{alg:sym_encoder_setProperty}.
We send any portions of the program trace involving symbolic objects to these
intermediate encoders.
The distinction between known and unknown properties is core to our symbolic
object encoding, with the symbolic object keeping track of any properties that
it has encountered before.
Each symbolic object is created with an initial map of known properties.
A \lstinline{setProperty} operation with a concrete property name marks that
property as known, and it will then on return the supplied value to preserve
JavaScript semantics.
The complementary \lstinline{getProperty} operation on a fixed property has normal
behavior, returning the (potentially symbolic) known value from the object.
So far, this encoding is straightforward and preserves standards semantics,
returning known property values for an object.
However, in order to explore the program symbolically, we need a special approach to treating
unknown property lookups. Whenever a program performs a \lstinline{getProperty}
on an unknown property we return a new, untyped, symbolic variable rather than
\lstinline|undefined| (the standard behavior). The specified
property of the symbolic object is then marked as known and fixed to this new
symbolic value. When a test case terminates, new tests will be created to
explore the program for each supported symbolic type.

There are a number of advanced features which can change
the behavior of \lstinline|getProperty| and \lstinline|setProperty| operations, such
as \lstinline|defineProperty|, which can trigger the execution of a function
instead of map lookup.
Methods can also be used to change the enumerability of properties within
an object.
We concertize the symbolic objects when handling these cases, and so our
encoding is under-approximate when modeling these behaviors.

\subsection{Mixed Type Arrays}
\label{sec:mixed_type_arrays}

We have described an approach to model objects, which are in
essence maps between string property names and values of any type.
We now show the same approach can be applied to arrays as well.
Conceptually, arrays are very similar to objects, mapping integer or string
property names to values.
The most significant differences between arrays and objects are the custom
behaviors of the length property, enumeration, and accompanying methods (e.g.,
\lstinline|push| and \lstinline|pop|).
In JavaScript, it is valid to also write to non-integer properties to arrays,
with the array acting as a object in these cases.
For example, \lstinline|let arr = [1,2,3]; arr['dst'] = '/home';| would yield \lstinline|[0: 1, 1: 2, 2: 3, length: 3, dst: 'home']|.

\begin{figure}
\includegraphics[scale=0.25]{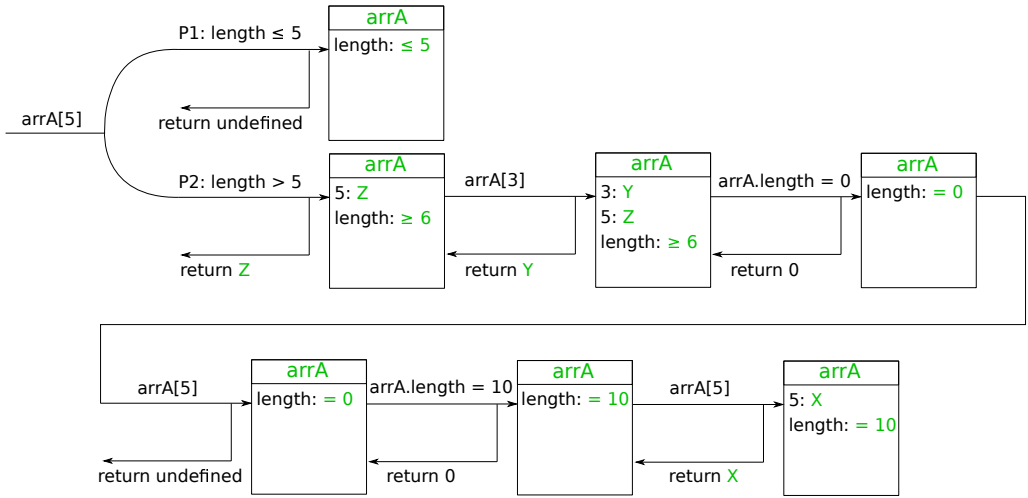}
\caption{Illustration of symbolic array modeling.}
\label{fig:array_expansion_explained}
\end{figure}

We intercept reads and writes to array length, which is a reserved property
name in arrays.
The array length property will always be one higher than the largest element
index in the array. This point is important because arrays do
not need to be contiguous (i.e., there may be gaps between two indices).
This design choice has an impact on enumeration, where looping on the array
length will include all indices \lstinline|0 <= index < arrayLength|, but using
the \lstinline|of| or \lstinline|in| operators will only include those which have
been set, since these operators will only include properties which are marked
as enumerable. For example, examine the following program:
\begin{lstlisting}
let arr = []
arr[0] = 1;
arr[4] = 2;
\end{lstlisting}
Here, the interpreter will yield the array \lstinline|[0: 1, 4: 2, length: 5]|. If we enumerate using the \lstinline|of| or \lstinline|in| operators we would see \lstinline|1| and \lstinline|2| enumerated upon, however if we enumerate and print all properties through the array length then we would see \lstinline{1, undefined, undefined, 2} printed.

When a program writes to the array length property, the array will be truncated or expanded to the new length.
If the value is less than the current array length, any values in the indices \lstinline|newLength <= index < oldLength| will be deleted from the array. If the value is greater than the current array length, then the array will be extended with \lstinline|undefined| values. To illustrate this, see the following program:
\begin{lstlisting}
let x1 = [1,2,3]
let x2 = [1,2,3]
x1.length = 100;
x2.length = 0;
\end{lstlisting}
In this example the variable x1 would have a length of 100 with all values
after 3 being \lstinline|undefined|, while x2 will be empty.

We illustrate these changes in behavior through
\autoref{fig:array_expansion_explained}. To ensure we accurately model array
length, we create a separate symbolic integer to represent it. This value is
initially unbounded and has constraints applied as the program executes. As
we fetch property 5, we explore two paths, one where the existing length
property is large enough to accommodate the new value and one where it is
not. In the case where it is not the value of the property will be undefined,
and in the other case it will return a new symbol using the same approach
as our symbolic objects. The second step illustrates what happens when an
array lookup occurs on an array that is longer than our property index. Here,
a second path is infeasible because the array length cannot be less than
six. Direct writes to an array fix the symbolic length; writing a length
of zero to the array truncates it, removing all properties. Subsequent property
lookups will all return undefined. A write of length 100 expands the array
to a fixed length but does not fix any properties. Here, a property lookup creates
a fresh symbol because the previous one was erased. The new symbol is given
a unique name in the path condition, and can interact with the symbol that
used to occupy this property.

\subsection{Optimized Support for Homogeneously Typed Arrays}
\label{sec:hom_arrays}

The final component of our encoding is a direct translation to SMT for
homogeneously typed arrays.
This encoding enables symbolic property names in homogeneously typed
arrays. As motivated previously, directly encoding JavaScript arrays in
SMT is too expensive for DSE, since we would need to encode potentially
recursive values in SMT. Our generic array and object encoding overcome
this by on-demand symbol generation, but this strategy cannot reason about
property indices symbolically. For example, in the following program, we
will not exercise the error:
\begin{lstlisting}
let i = I with initial vale 0;
let arr = A with initial value [];
 if (arr[0] == 5 && arr[i] != 5) {
 throw 'Error';
}
\end{lstlisting}
In this program, we do not exercise the error because we concretize symbolic
property names. Thus, \lstinline|arr[i]| will resolve to \lstinline|arr[0]|,
leading to an infeasible constraint of $arr[0] = 5 \land arr[0] \neq 5$,
and will not consider any paths where i is not 0 due to concretization. If we
set \lstinline|i| to \lstinline|1|, then this error would be found. We provide an
encoding for homogeneously typed arrays directly in SMT to explore portions
of a program where property name concretization is limiting analysis. Since the
encoding is directly in SMT, we no longer need to concretize property names,
allowing us to reason about property names symbolically.

\begin{algorithm}[t]
 \If{$0 \leq index < base.length$}{
  $PC \leftarrow PC \land 0 \leq index < base.length$\;
  \Return{select(base, index)};
 } \Else {
  $PC \leftarrow PC \land (index < 0 \lor index > base.length)$\; 
  \Return{undefined}\;
 }
 \caption{Homogeneous Array -- getProperty(base, index).}
 \label{alg:homog_getProperty}
\end{algorithm}

\begin{algorithm}[t]
 \If{$index > 0$} {
  base.length = $index + 1$ if $index \geq base.length$ otherwise $base.length$\;
  base = $store(base, index, value)$\;
 }
 \Return{value}
 \caption{Homogeneous Array -- setProperty(base, index, value).}
 \label{alg:homog_setProperty}
\end{algorithm}

\begin{algorithm}[t]
 knownValues = []\;
 \For{i in Concrete(base.length)} {
  knownValues[i] = $select(base, i)$\;
 }
 \Return{GenericArray(knownValues, base.length)}
 \caption{Homogeneous Array -- downgrade(array)}
 \label{alg:homog_downgrade}
\end{algorithm}

Our encoding uses existing SMT solver support to
represent arrays.
A typed array has two symbolic components, the array data
and array length.
The array base is a symbolic mapping of integer property names to symbolic
values of the array's type.
The symbolic length property is used to represent the current constraints on
array length, which is necessary to test out-of-bounds array element access.
A symbolic \lstinline|getProperty| can explore two paths, one where the array
is shorter than the index resulting in \lstinline|undefined|, the second
where the array includes the index, resulting in a value of the array
type. \lstinline|setProperty| operations update the symbolic length to accommodate
the new value and then inserts it into the base. This is illustrated in
\autoref{alg:homog_getProperty} and \autoref{alg:homog_setProperty}. In these
algorithms, the methods \lstinline|select| and \lstinline|store| map directly to SMT.
We downgrade when a \lstinline|setProperty| is given a value that is not the array
base type.

The process for downgrading a homogeneously typed array to a mixed-type array
is detailed in \autoref{alg:homog_downgrade}.
Array downgrading converts a homogeneously typed array into a generic array
to allow mixed types.
We do this by using the concrete array length to derive the initial mapping
for the mixed-type array.
We copy the homogeneously typed array's length into the new array so that we
respect existing length constraints.
 
\section{Evaluation}
\label{sec:eval}

We now set out to show the effectiveness of our approach on a subset of
JavaScript built-in functions introduced with the ES6 specification. Here,
we set out to answer the following research questions:
\begin{enumerate}[start=1,label=\bf RQ\arabic*:,leftmargin=\widthof{\textbf{RQ1:}}+\labelsep+2em]
  \item \label{item:conformance_rq1} Is our approach able to cover the logic of built-in functions?
  \item \label{item:conformance_rq2} Can our approach find any bugs in built-in methods?
  \item \label{item:conformance_rq3} Does the addition of our test cases improve coverage of Test262?
\end{enumerate}

We answer these research questions through three experiments on selected
functions introduced with the ES6 specification.
In the first experiment we evaluate the effectiveness of our conformance
test case generation strategy using the polyfills \lstinline|core-js| v3.1.4 
and \lstinline|mdn-polyfills| v5.17.1.
Here, we show that ExpoSE achieves high coverage of many method
implementations.
For our second experiment we use our generated conformance test suite and
voting mechanism to search for errors in existing implementations of the
ES6 standard, finding 17 bugs in a widely depended built-in
implementation.
Finally, we evaluate the coverage of our test suite against Test262
under the QuickJS interpreter (v2019-10-27). In this study we see that, while
Test262 generally covers more branches of tested methods overall, our test cases explore parts of the
built-in implementations which are not covered by Test262.

\subsection{Test Case Generation}
\label{sec:polyfill_testcase_generation}

In our first experiment we answer \hyperref[item:conformance_rq1]{RQ1} through
an evaluation on two popular ES6 built-in method implementations
found on NPM. We extracted our surrogate implementations from \lstinline|core-js| 
and \lstinline|mdn-polyfills|. Overall, we collected 96,470 new
unique test cases. We show that we achieve high coverage of the built-in
implementations during symbolic execution, suggesting that a large portion
of the implementation is covered.

\subsubsection{Methodology}

Our test harness loads the portions of the library we wish to test and selects
a target method. The method is then executed with symbolic arguments for both
the \lstinline|this| argument and each of the method arguments. We analyze this
harness with ExpoSE. Each method is tested in isolation, through a single
analysis using ExpoSE with a timeout of one hour on a 64 core machine. After
analysis, the generated test cases are combined and duplicates are removed.

\subsubsection{Results}

\begin{table}[t]
\begin{tabularx}{\columnwidth}{X|r|r|r}
\toprule
\bf Function & \bf Test Cases & \bf core-js Coverage & \bf mdn-polyfill Coverage \\
\midrule
Array.from & 13122 & 90\% & 84\% \\
Array.of & 162 & 84\% & 82\% \\
Array.fill & 2645 & 89\% & 85\% \\
Array.filter & 81 & 88\% & N/A \\
Array.findIndex & 162 & 72\% & 51\% \\
Array.forEach & 81 & 78\% & N/A \\
Array.reduce & 729 & 76\% & N/A \\
Array.some & 162 & 84\% & 35\% \\
String.endsWith & 64179 & 86\% & 82\% \\
String.includes & 15957 & 93\% & 91\% \\
String.padStart & 13220 & 94\% & 94\% \\
String.padEnd & 13220 & 94\% & 94\% \\
String.repeat & 2066 & 96\% & 83\% \\
String.startsWith & 4215 & 91\% & 88\% \\
String.trim & 2025 & 95\% & 83\% \\
\bottomrule
\end{tabularx}
\caption{Automatically generated test cases by built-in method.}
\label{tbl:automatically_generated_cases}
\end{table}

We generated 129,960 new test cases overall, which was reduced to 96,470
after removal of duplicate tests. \autoref{tbl:automatically_generated_cases}
presents the results of our evaluation, providing coverage information from
the analysis of the \lstinline|core-js| and \lstinline|mdn-polyfills| variant if the method
was supported by that library. Overall, we found that our prototype is more capable
of generating test cases for string methods than array methods.
These results are inline with our expectations, as the string support in
ExpoSE is mature. Further improvements in ExpoSE modeling and SMT solvers
could improve this support even further. In particular, our encoding currently
does not include symbolic models for array methods other than \lstinline|push|,
\lstinline|pop|, \lstinline|includes|, \lstinline|indexOf|, which may lower overall
performance.

\subsection{Executing Our Test Cases}

We have now generated a suite of test cases for our selected methods
and are ready to test built-ins. In this section we set out to answer
\hyperref[item:conformance_rq2]{RQ2} by executing our tests on five JavaScript
built-in implementations.
We execute each of our generated test cases on three interpreters
and two polyfill implementations.
To analyze the output of these test cases, we construct the voting mechanism
outlined in \autoref{ssec:test_executor} from our selected interpreters. Each
test case is executed once per interpreter, and after they finish they vote
on the correct output. We found 17 unique bugs in \lstinline|mdn-polyfills|,
showing that our approach is effective in generating useful test cases for
conformance testing. We did not find bugs in any other implementations but
this was expected as the methods tested are from a mature standard.
We found zero cases which required manual intervention during voting.

\subsubsection{Methodology}

We selected QuickJS 2019-10-27, SpiderMonkey 68 (through the standalone interpreter), Node.js v8.12.0,
\lstinline|core-js| v3.1.4 and \lstinline|mdn-polyfills| v5.17.1 for testing. We tested each of the test-cases identified in
\autoref{sec:polyfill_testcase_generation}. We executed each test case once
with each competing implementation and stored the output. Next, we examined the
result of each test case for divergence between the tested implementations. If
there is any divergence then we used the outlined voting mechanism to resolve
the failing case. Test cases were each executed with a maximum time of 10
minutes on each interpreter, though no test cases hit this boundary. Tests
which crashed or exceeded the timeout are terminated with a failure.

\subsubsection{Results}

\begin{table}[t]
\begin{tabularx}{\columnwidth}{X|r|r|r}
\toprule
\bf Implementation & \bf Unique Exceptions & \bf Test Case Failure & \bf Bugs \\
\midrule
mdn-polyfills~\cite{MDNPolyfill} & 34 & 200 & 17 \\
core-js~\cite{core-js} & 63 & 125 & 0 \\
SpiderMonkey~\cite{SpiderMonkey} & 72 & 66 & 0 \\
Node.js~\cite{NodeJS} & 56 & 122 & 0 \\
QuickJS~\cite{QuickJS} & 24 & 141 & 0 \\
\bottomrule
\end{tabularx}
\caption{Test case summaries for 5 built-in implementations.}
\label{tbl:exceptions_and_faults}
\end{table}

\autoref{tbl:exceptions_and_faults} presents a summary of test case executions
for the 5 built-in implementations. \textbf{Unique Exceptions}
gives the number of unique exceptions identified across the executions of
all test cases (i.e, where an exception text has not been seen before after
test specific details are removed). \textbf{Test Case Failure} details the
total number of test cases where the interpreter failed to give a result
due to crash or timeout. The final column, \textbf{Bugs} gives the number
of bugs found in each implementation.

Our test case executor found 17 bugs automatically, all within \lstinline|mdn-polyfills|. We were able to confirm these bugs through manual
analysis.
For example, in one test case we observed that
\begin{sloppy}\lstinline[breaklines=true]|String.prototype.includes.apply([0,0], [[]])|\end{sloppy} should yield \lstinline|true|, but in \lstinline|mdn-polyfills|
the built-in returns \lstinline|false|. We found this divergence occurs
because the implementation does not coerce the \lstinline|[0, 0]| to a string.
The identified bugs show that a consensus based test executor can
be used to verify the correct behavior of built-in JavaScript methods.
Manual analysis found that the bugs identified were all triggered by
unconsidered type coercions in string and array methods. In some cases,
this led to the method producing output when it should have thrown an error.
In others, the method produced an incorrect output, such as
\begin{sloppy}\lstinline[breaklines=true]|Array.includes|\end{sloppy}, which
would return \lstinline|true| when it should have returned \lstinline|false|
on some inputs.

In addition to finding some bugs, we exercised many unique exceptions in
interpreters. The high number of unique exceptions suggests that our test suite
is exploring many interesting corner cases of implementation. Interestingly, we
do not see the same number of unique exceptions across interpreters. We found
that some implementations have much more verbose error messages
for built-ins than others. While the exception messages are not standardized,
and so this is not an implementation error, the lack of verbosity could make
errors harder to debug. 

We experienced some test case failure for each of the implementations
tested. We observed zero cases of failure due to test timeouts or interpreter
error; instead, all observed failures were due to interpreter memory
limits. Most of these errors occur in \lstinline|String.repeat|, where many
of the test inputs are large values which hit interpreter memory limits. We
examined our surrogates to understand why the DSE engine is generating such
extreme cases. We find that in one of our surrogate implementations there
is an upper limit on string size through a boundary condition \lstinline|if (str.length * count >= 1 << 28)|. The condition drives ExpoSE to generate
a series of test cases supplying large arrays or strings as input. The
specification does not specify interpreter memory limits, so the different
number of failing cases is not an error. In particular, we observed that SpiderMonkey
avoids test case failure in these cases by having stricter limits on bounds
for \lstinline|repeat|. As an example, at the time of writing, Node.js
will execute \lstinline|'h'.repeat(1 << 28)| but SpiderMonkey will not. In
the specification, ECMAScript does not add any constraints to the range of
strings, so long as they are positive integers. In practice, the reason we
see these memory errors in QuickJS and Node, but not SpiderMonkey, is because
string boundaries are explicit in SpiderMonkey method implementations. So
these errors manifest as exceptions without crashing the interpreter.

Our study has shown that we can detect faults in a real built-in implementation
with 35,000 weekly downloads at time of writing. The ability to detect real
bugs using our approach shows that a consensus based approach for test case
evaluation can be effective. In addition, our approach generated a large
number of unique exceptions in the tested cases and covered an obscure
difference in string length constraints between interpreters, demonstrating
that our test cases explore interesting paths through the implementations.

\subsection{Test Suite Coverage}

To ensure that our new approach generates novel test cases, we now compare the
branch coverage of the new test cases to Test262. We show that the
addition of our test cases leads to an increases in overall branch coverage
in QuickJS, demonstrating that our approach is generating novel test cases.
 
To test our approach we built a version of the QuickJS interpreter with support for gcov so we could collect internal code coverage metrics.
QuickJS is a complete ES6 implementation of JavaScript~\cite{QuickJS}.
We selected this interpreter because it executes the code in a purely
interpreted manner, without JIT or other runtime optimization, and it has
built-ins implemented directly in its source code.
This is important as many prominent engines, including Node.js~\cite{NodeJS} and
SpiderMonkey~\cite{SpiderMonkey}, do not implement language built-ins directly in source code.
Instead, these engines implement a small subset of JavaScript in their native
language and then implement the remaining built-ins in JavaScript.
Implementing built-ins in JavaScript allows these engines to take advantage
of JIT optimizations and reduce engine development time, but, this makes it
challenging to collect coverage metrics as built-in functions do not have
a clear instrumentation point.

In our study, we found that our automatically generated conformance test suite improves branch coverage by up to 15\%. We see coverage improvements in almost every tested function, demonstrating that the approach is versatile. Our results show that we can use automatically generated test cases to supplement the Test262 suite to provide greater over-all coverage of JavaScript interpreters.

\subsubsection{Methodology}

We modified the QuickJS build process to include support for branch coverage
output via gcov, a tool which collects coverage information through compile
time instrumentation. For each built-in method, we then executed all of
our generated test cases and the relevant portion of the Test262
suite. Once each had finished, we extracted the covered branches, using a
manual analysis to identify the appropriate function names in the QuickJS
source code.

When evaluating the coverage of a function within a program, we present both
shallow and deep metrics for combined coverage increases, and follow calls
to a depth of 3 when presenting absolute branch coverage.
If we only present shallow coverage metrics (i.e., we do not follow function
calls), then we may under-represent coverage improvements as logic for
built-ins is often spread across many methods.
Conversely, including all reachable functions may make our results less
insightful by including large amounts of indirectly related code, such as utility methods, which may also be called by other methods during execution.
By presenting our combined coverage improvements at different call depths,
the reader can see how branch coverage changes as we follow an implementation
deeper into the methods it calls.

In our coverage metrics we only include methods defined in the core QuickJS
implementation and do not include library calls.

\subsubsection{Results}

\begin{table}[t]
\begin{tabularx}{\columnwidth}{X|r|rr|rr|rr}
\toprule
\bf Function & \bf Total Branches & \bf ExpoSE & \% & \bf Test262 & \% & \bf Combined & \% \\
\midrule

array\_from & 1075 & 640 & 60\% & 802 & 75\% & 957 & 89\% \\
typed\_array\_from & 897 & 572 & 64\% & 696 & 78\% & 829 & 92\% \\
array\_of & 875 & 509 & 58\% & 640 & 73\% & 754 & 86\% \\
typed\_array\_of & 512 & 285 & 56\% & 405 & 79\% & 457 & 89\% \\
array\_fill & 740 & 436 & 59\% & 586 & 79\% & 663 & 90\% \\
typed\_array\_fill & 222 & 165 & 74\% & 180 & 81\% & 206 & 93\% \\
array\_every & 953 & 564 & 59\% & 748 & 78\% & 873 & 92\% \\
array\_find & 527 & 294 & 56\% & 422 & 80\% & 471 & 89\% \\
typed\_array\_find & 1472 & 869 & 59\% & 1103 & 75\% & 1302 & 88\% \\
array\_reduce & 702 & 404 & 58\% & 556 & 79\% & 636 & 91\% \\
array\_includes & 734 & 430 & 59\% & 589 & 80\% & 667 & 91\% \\
string\_includes & 220 & 148 & 67\% & 183 & 83\% & 194 & 88\% \\
string\_pad & 139 & 85 & 61\% & 112 & 81\% & 121 & 87\% \\
string\_trim & 54 & 34 & 63\% & 46 & 85\% & 47 & 87\% \\
string\_repeat & 139 & 89 & 64\% & 112 & 81\% & 116 & 83\% \\

\bottomrule
\end{tabularx}
\caption{Branch coverage for systematically generated conformance tests and Test262 at a call depth of 3. }
\label{tbl:overall_comparison_conformance}
\end{table}

\autoref{tbl:overall_comparison_conformance} details total number of branches,
branches covered by our systematically generated conformance tests (ExpoSE),
and branches covered by Test262.
We selected a call depth of 3 as following calls further included many
utility methods, making results less insightful.
Here, tests generated by our approach achieve reasonable branch coverage,
but do not exceed the coverage of Test262 which is already very
high for every method.
When we combine the branches covered by automatically generated conformance
tests and Test262 we see an overall coverage improvement over
Test262 for every tested function, demonstrating that generated conformance tests are exploring new routes through the implementation.

\autoref{tbl:quickjs_coverage_improvement} shows the results of our coverage
study at various call depths.
The function names in the table are the internal function names in
QuickJS. QuickJS sometimes implements optimized methods for typed arrays,
which is why there may be two methods for the same feature.
We see branch coverage improvements in many of the methods we test, in some cases seeing a 15\% improvement overall. Our results demonstrate that our automatically generated test cases do explore further into built-in method behavior than Test262 answering \hyperref[item:conformance_rq3]{RQ3}.
In most functions, we see notable coverage increases, even at a call depth
of 0 (i.e., not including the coverage impact of any called methods).
These results highlight that our approach is exploring untraveled paths
through built-in function implementations login, and not just expanding
coverage in utility methods.
The coverage increases at low call depths show that built-in specific
edge cases are being exercised, as these expressed near the surface of
the call-tree.

Our study of interpreter coverage achieved between automatically generated conformance tests and Test262 shows that supplementing Test262 with automatically generated test cases will improve the test suite. We found that our approach would improve branch coverage of the test suite by up to 15\% in a complete ES6 JavaScript engine. These improvements demonstrate that our method can improve conformance testing for JavaScript interpreters using only automatically generated test cases. Such coverage improvements in the testing suite raise the likelihood that implementation errors will be detected before they cause problems in the wild.

\begin{table}[t]
\begin{tabularx}{\columnwidth}{X|r|r|r}
\toprule
\bf Function & \multicolumn{3}{c}{\bf +Branches\% (Depth)} \\
& 0 & 3 & 5 \\
\midrule
array\_from         & +4.76\% & +14.42\% & +14.68\% \\
typed\_array\_from   & +6.14\% & +14.82\% & +13.48\% \\
array\_of           & +0\% & +13.03\% & +12.83\% \\
typed\_array\_of     & +41.67\% & +10.16\% & +14.19\% \\
array\_fill         & +0\% & +10.41\% & +10.57\% \\
typed\_array\_fill   & +8.33\% & +11.70\% & +14.97\% \\
array\_every        & +8.33\% & +13.12\% & +12.64\% \\
array\_find         & +8.51\% & +9.3\% & +11.41\% \\
typed\_array\_find   & +11.43\% & +13.52\% & +15.32\% \\
array\_reduce       & +3.13\% & +11.4\% & +12.36\% \\
array\_includes     & +1.67\% & +10.63\% & +11.76\% \\
string\_includes    & +0\% & +5\% & +9.69\% \\
string\_pad         & +0\% & +6.47\% & +10.68\% \\
string\_trim        & +0\% & +1.85\% & +8.37\% \\
string\_repeat      & +0\% & +2.88\% & +10.52\% \\
\bottomrule
\end{tabularx}
\caption{Coverage improvements of automatically generated tests at various call depths by built-in method implementation.}
\label{tbl:quickjs_coverage_improvement}
\end{table}

\section{Related Work}

We now briefly review related work in the space of dynamic symbolic execution, with a particular emphasis on memory models and handling of symbolic reads and writes to objects and arrays.

Mayhem~\cite{mayhem_paper} is a dynamic symbolic execution engine for compiled
programs that represents a 32-bit address space symbolically to model
program memory.
In this work a symbolic memory model improved the effectiveness of DSE by 40\%, showing that supporting symbolic memory is crucial.
To make their solution feasible, they limit the symbolic representation to
reads and do not consider writes symbolically.
EXE~\cite{exe} supports a single-object model, where pointers are
concretized and only a single address is considered.
The approaches are similar to our own when treating symbolic field names,
where ExpoSE concretizes the field name to avoid exploring an unbounded
number of inputs.

S2E~\cite{ChipounovKC11} models system memory symbolically in a symbolic
machine emulation, achieved through instrumentation of memory reads and writes.
Modeling memory interactions in low level applications is very different
from JavaScript, since memory is fixed type and the DSE engine does not need
specific encodings for language structures.

The DSE engine KLEE~\cite{klee} supports multiple memory models, including a
forking approach where one path is created to explore each symbolic memory
region, and a flat approach which reasons about memory as a single continuous
block.
Recent approaches split memory regions into segments
to allow more efficient analysis~\cite{segmented_memory_klee}.
These approaches are highly tailored to reasoning about systems memory with C
style pointers and are not directly applicable to JavaScript object modelling.
 
Bucur et al.~\cite{chef-asplos14}, found that useful symbolic execution of
JavaScript interpreters and their programs is out of reach of current binary
symbolic execution engines.
This work highlights the need for symbolic
dynamic language interpreters, where knowledge of the language structure
can make symbolic execution feasible.

There has been work to enable automated testing for Java~\cite{HavelundP00, bogor, AnandPV07}.
Symbolic encodings for Java classes are insufficient
for JavaScript as they rely upon a known structures and
typing~\cite{generalized_symbolic_execution_for_model_checking,
test_input_generation_jpf, symstra}.
Our approach is similar to previous symbolic representations of maps, but
does not require fixed type fields.

\citet{type_tests_typescript} use TypeScript type specifications and feedback
directed random fuzzing to identify mismatches between type specifications
and observed behaviors.
Through this approach the authors identify many inconsistencies, motivating
the use of dynamic analysis for specification testing.

\citet{zesti} symbolically execute test suites to find bugs.
A symbolic execution runs on the existing harnesses used by a program for
unit testing, replacing concrete values with symbolic ones in order take
advantage of interesting test conditions.
Unlike our approach, only simple error conditions are considered because
the tool cannot deduce the expected output after a charge in input.

\citet{shadow_of_a_doubt} use DSE to automatically discover differences in
behavior between program versions.
The authors test versions of the same software, while our approach
tests differences between many implementations of the same specification.
As versions of the same software are tested but program specifications are
not static, it is difficult to decide whether changes in behavior between
two versions are desired.
This differs from our approach, where the behavior of compliant
implementations is fixed and divergence is an error.

\section{Conclusions}

We have presented a new approach to automated generation of conformance tests
for the ECMAScript language specification based on dynamic symbolic execution of
polyfills. To adapt symbolic execution to this setting, we introduced a new
model for generating structured inputs in the presence of dynamic types. 
We evaluate our method on selected functions from JavaScript built-in
implementations, generating 96,470 new conformance test cases from two packages
of polyfills. Using majority voting in place of usual test oracles, we found 17
pre-existing bugs in JavaScript built-in implementations. Our new test cases
improve branch coverage of the Test262 implementation conformance test suite by
up to 15\%.

Overall, our approach promises to make JavaScript conformance testing more
thorough and simpler to set up in the future. Given that often polyfills are
written before standardization to experiment with new language features, our
method can derive corresponding conformance tests directly from these
implementations.


\end{document}